# Test Plan Generation for Live Testing of Cloud Services

Oussama Jebbar
*Gina Cody School of Engineering and Computer Science*
*Concordia University*
Montreal, Canada
ojebbar@encs.concordia.ca

Ferhat Khendek
*Gina Cody School of Engineering and Computer Science*
*Concordia University*
Montreal, Canada
ferhat.khendek@concordia.ca

Maria Toeroe
*Ericsson Canada Inc.*
Montreal, Canada
maria.toeroe@ericsson.com

*Abstract*— Live testing is performed in the production environment ideally without causing unacceptable disturbance to the production traffic. Thus, test activities have to be orchestrated properly to avoid interferences with the production traffic. A test plan is the road map that specifies how the test activities need to be orchestrated. Developing a test plan includes tasks such as test configuration selection/generation, test configuration deployment planning, creating the test runs schedule, choosing strategies to mitigate the risk of interferences, etc. The manual design of a test plan is tedious and error prone. This task becomes harder especially when the systems are large and complex. In this paper we propose an approach for automating test plans generation. With this approach we aim at reducing service disruption that may be induced by the testing activities in production. We illustrate our approach with a case study and discuss its different aspects.

## I. Introduction

Testing in the production environment is becoming a necessity. It is used for several purposes such as service composition [2], fault localization [3], evaluation of business objectives [4], etc. Needless to say that such testing needs to be conducted without unacceptable disturbance to the production traffic. Testing a system in its production environment without causing unacceptable disturbance is called live testing [1].

The main challenge of live testing is to avoid test interference, which is an alteration, degradation, or loss of a system's property due to the coexistence of test activities and production traffic. In other words, the coexistence of the test activities and production traffic leads to a violation of one of the system's functional or non-functional requirements. The countermeasures taken to alleviate the risk associated with test interferences are known as test isolation [3]. Other challenges such as diversity of test cases and their runtime environments, short reaction times due to stringent non-functional requirements such as high availability, the number of test configurations under which test cases are to be run, etc., add to the complexity of conducting testing activities in modern production environments such as clouds and zero touch networks. Because of this increased complexity, handling manually testing activities such as test planning becomes tedious and error prone.

Test planning is the process of developing a test plan [11]. Test management decisions such as the determination of test cases schedule, test configurations selection, determination of the resources required for test execution, etc. are made during test planning. Furthermore, planning for live testing requires making some extra decisions such as the selection of the isolation countermeasures. In this paper, we address the challenges related to test planning for live testing. We propose an approach for the automated generation of test plans. The generated test plan is described using the UML Testing Profile (UTP) [5], and it is based on the architectural and modeling framework proposed in [1]. The test methods used to run the test cases in production have been proposed in [13]. The main goal of our test plan generation method is to guarantee that at the execution of the generated test plan any disturbance remains within an acceptable range. The method is based on the observation that all the test plans that execute the same set of test cases, each of which is mapped to a set of test configurations under which it should be run, have the same number of test runs, i.e. the same cost (time wise and disturbance wise) for running the test cases. Therefore, the only aspect by which one test plan outperforms another one is in the handling of the deployments of test configurations.

The rest of this paper is organized as follows. In Section II we review the related work. We provide some background in Section III. In Section IV we discuss the modeling using UTP of the test methods introduced in [13]. In Section V we introduce and discuss our test plan generation method. We present an illustrative example in Section VI before concluding in Section VII. The proofs and evaluation of different activities composing our approach can be found in the appendices.

## II. Related work

The concept of test plan has more than one definition in the literature. As a result, the concerns of test plan generation methods in the state of the art depend heavily on the used definition. The work in [6] for instance uses "test plan" to refer to the test suite. It proposes a method for test suite generation, more specifically test suite generation for GUI applications. A broader definition of test plan is used in [9], it is seen as an artifact that documents the test scope, test configurations, and test cases used to validate a new version of a product. [9] thus describes a CASE tool that can be used to generate a test plan to document testing of a new version of a product from the test plans that document testing of older versions of that same product.

ISO29119-1 [10] defines a test plan as a detailed description of test objectives to be achieved as well as the means and schedule for achieving them, organized to coordinate testing activities for some test item or a set of test items. Furthermore, ISO29119-2 [11] defines the test planning process as the process used to develop a test plan. Test planning according to this standard consists of several steps among which we find, identification and analysis of risks, identification of risk mitigation approaches, designing test strategy, and determining the test schedule. The definitions we use coincide to some extent with the definitions proposed in the ISO standard. In fact, the test plan as we propose it includes the test objective (of the test session) as well as the means to achieve it (the test cases and test configurations). Furthermore, the test plan generation approach that we propose covers creating the test schedule as well as identification of risks (applicability check of test methods) and risk mitigation approaches (the test methods we use in our approach [13]) and the design of test strategy (test method selection). [8] also covers a fair share of these steps as it proposes a method that starts from user configured test parameters, which include at least one test objective, to generate a test execution plan. The test execution plan is a set of actions that achieve the test objective, and that are generated by applying a set of rules (derived from the user provided input and some pre-set rule templates). This method does not explicitly handle the creation of the test schedule, although it can be adapted to achieve that. However, it can select the test actions and determine the test resources (tester, runtime environment, etc.) needed to execute them.

[7] addresses the test resource allocation problem by estimating the duration (man hour) needed to test a new product. When such a duration is fixed as a budgetary constraint for instance, [7] generates an estimation of the risk associated with testing for that duration based on the defects that will be detected. Only a few test planning methods proposed in the literature design test strategies to mitigate the risk of interferences. The work in [3], for example, proposes a set of algorithms to select test isolation countermeasures that will minimize the cost of implementation. This selection is also balanced with the cost of testing, i.e. how early defects are detected, and the cost of diagnosis, i.e. accuracy of the fault localization. The work in [12] also handles the risk of interferences using the same test isolation methods as [3]. In addition, it also addresses the resource allocation concern of test planning by minimizing the resource consumption impact of testing in the production system.

### III. Background

To reduce the cost of their services, cloud service providers satisfy the requirements of their tenants using configurable software which can be configured differently to provide different features with different characteristics. Configurations can be tenant configurations, application configurations, or deployment configurations.

#### A. Configured Instance and Service Instance

Applying a set of configurations to a configurable software yields a configured instance (CI). The workload handled by a single CI is called a service instance (SI). The requirements of a tenant are satisfied using a service which consists of one or multiple SIs.

Configurations play various roles in the behavior and operation of configurable software. Application configurations are used to expose/refine features of the configurable software which are parameterized differently for different tenants using tenant configurations. When instantiated, a CI yields a set of components, each on a separate node, which are actively providing the actual SI. The number of such components, their locations (physical or virtual nodes that may change over time), their interactions with components of other CIs, the policies that govern the number of such components, etc. are aspects set using deployment configurations. Furthermore, the scalability of the CIs is also set at deployment configuration level. Parameters that are used to configure scalability include the cool down period and scaling step. The cool down is the minimum period between two consecutive scaling actions. The scaling step sets the number of components instantiated (resp. terminated) in each scaling out (resp. scaling in) action.

The number of components of each CI, their locations, and their binding information change over time due to recoveries from failures as well as scaling actions. Such information is captured in the runtime configuration state of a system. The set of runtime configuration states a system can enter depends on the system's configuration. When the system is in a given runtime configuration state, each component is located on a specific node, in a specific network, sharing that node with a set of components from other CIs. The location information (node and network) and collocation information define the environment under which the component is currently serving. Therefore, a runtime configuration state is identified by the set of environments under which the components of the CIs are serving when the system is in that runtime configuration state. Furthermore, we can also identify runtime configuration states by the environments under which the SIs that compose each service are provided. For each service, such a combination of environments is called the path through which the service is provided. Note that for services that are composed of a single SI the concept of path coincides with the concept of environment as there are no combinations of environments to consider at this level. As a result, the concept of path, as we define it, is not to be confused with a path in white box testing which may refer to control flow path or data flow path. To validate the compliance of services to their requirements, cloud service providers use test cases. These test cases may involve one or more CIs depending on the requirements the test case covers.

#### B. Modeling a test plan using UML Testing Profile

The modeling of a test plan using UTP is done through the mapping between the concepts we proposed in [1] and the concepts defined in UTP as shown in Table I. This mapping models a test plan as a TestExecutionSchedule that runs UTP TestCases. UTP TestCases consist of one or more test cases provided by the vendor or the developer along with a test configuration. UTP TestProcedures are used to model invocations of vendor provided test cases and may be modeled using UML concepts. UTP also offers the possibility of specifying TestProcedures using other languages as OpaqueBehavior (a concept inherited from UML). UTP TestCases also include a setup ProcedureInvocation which is used for preparation and deployment of the test configuration, and a teardown ProcedureInvocation which is used to tear down the test configuration. Test configurations in UTP include modeling the configuration of the test component as well as the configuration of the test item (system or component under test). Our modeling framework is agnostic to the pattern with which these configurations are modeled (as a class or a constraint) although we recommend modeling these configurations as constraints.

TABLE I. MAPPING THE ARTIFACTS IN THE ABSTRACT ARCHITECTURE TO UTP CONCEPTS

| Abstract Architecture concepts | UTP concepts |
| --- | --- |
| Test suite item in the test plan | ProcedureInvocation in the main phase of a TestCase |
| Test suite item runs | TestCase |
| Test plan | TestExecutionSchedule |
| Test preparation including the setting up of isolation countermeasure | TestCase setup procedure invocation |
| Test completion including the cleanup of isolation countermeasure | TestCase teardown procedure invocation |
| Test goal | TestRequirement or TestObjective |

## IV. MODELLING THE TEST METHODS

In a previous work [13] we proposed a set of test methods that can be used to perform live testing of cloud services. These test methods are applicable in an environment that supports 1) snapshotting and cloning of components; and 2) service relocation as means of state transfer between components. The single step is a test method which can be used to test services for which there is no potential risk of interferences. Using the single step method, one can iteratively set up some paths to be tested, execute the test case on the paths that were set up, remove these paths, and then proceed to the next iteration until all the paths have been tested. The small flip test method is a test a method that can be used when there is a potential risk of interferences and the number of components, say K, needed to provide the SI is less than half the number of nodes on which the CI is deployed. Using a small flip, one proceeds in two iterations 1) in the first iteration the paths that are set up are the ones that involve K nodes not currently used by the CI to provide the SI; and, 2) the second iteration tests the paths that involve the rest of the nodes on which the CI is deployed. Note that between the first and the second iterations one has to relocate the SI to be provided through the paths that were tested in the first iteration. When the available resources do not allow to use a small flip, we proposed the use of the rolling paths test method. In a rolling paths test method, one iteratively sets up one path at a time, tests it, and removes it then moves forward to test the next path (following the same steps) in the next iteration until there is no path to test. In this case, going from one iteration to another often involves a service relocation, although usually not the entire SI at a time. Finally, when a service relocation induces intolerable disturbance and sufficient resources are available, we proposed the big flip test method which consists of creating a new CI, the test CI, which is tested using the single step test method, then the SI is relocated to the tested CI and the old CI is removed.

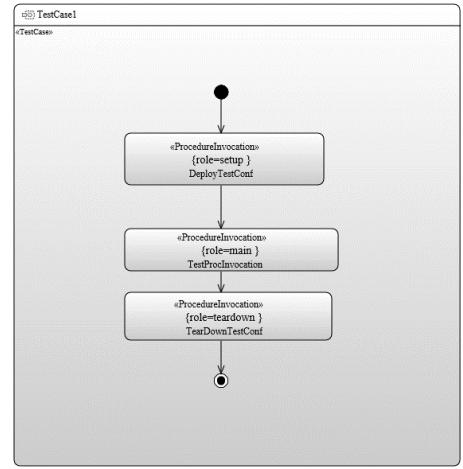

Fig. 1. Modeling TestCase with UTP

Test cases need to be run under paths (test configurations) that are representative of the runtime configuration states in which the system can be. These runtime configuration states are described by the environments of the components of the system when the system is in that runtime configuration state. We have proposed in [13] a set of coverage criteria that enable the tester to exercise a set of environments that is representative of the environments in which a component can be in the various runtime configuration states. Two important concepts to define such coverage criteria are the boundary environment and mixtures. A boundary environment is defined as the maximum collocation under which a component can serve in a given location, while a mixture is defined as an assignment of a number of occurrences over the set of boundary environments of a given CI. The sum of these assigned numbers of occurrences is called the mixture width. [13] describes a set of coverage criteria for mixture of width one (also known as paths), as well as coverage criteria for mixtures of arbitrary width.

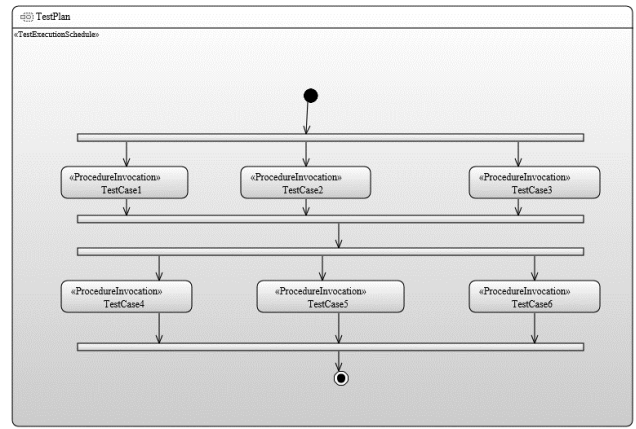

Fig. 2. Modeling of a single step test method

The test methods are patterns in which the test runs can be arranged to isolate the test traffic from the production traffic. Each UTP TestCase combines a set of one or more test suite item (TSI) runs that are run under the same TestConfiguration, i.e. the same path. Fig. 1 shows a TestCase with three invoked behaviors: 1) a behavior to deploy the test configuration, i.e. setup the path to be exercised, 2) a behavior that invokes the TSIs to execute against this path, and 3) a behavior to tear down the test configuration. Note that UTP offers roles that can be assigned to such invocations, namely, setup, main, or teardown.

Since the test methods iteratively execute the TSIs against one or more paths (in each iteration), until all the required paths are exercised; these test methods can be modeled as CombinedFragments of UTP TestCases. The specification of a CombinedFragment, per test method, goes as follows:

- Single step: each iteration of a single step test method is modeled as a ParallelFragment. UTP TestCases corresponding to the paths of an iteration are invoked in the fragment of that iteration. Fragments of the iterations are then put sequentially. Fig. 2 illustrates such pattern for a situation where a set of TSIs is to be executed under six paths and the maximum number of paths that can be deployed at once is three. Therefore, the model ends up with a first parallel fragment that executes the TSIs under the first three paths, and then deploys the next three paths to execute the TSIs against.

- Rolling paths: the rolling paths is modeled as a sequence of UTP TestCases Fig. 3 illustrates a rolling paths test method modeled in UTP. In the rolling paths, only a single path can be setup at a time, therefore the UTP TestCases are invoked sequentially. Each invoked UTP TestCase executes a set of TSIs against a test configuration (i.e. a path).

- Small flip: the small flip is modeled as two consecutive single steps (Fig. 2), targeting two disjoint sets of paths, separated by a service relocation ProcedureInvocation.

- Big flip: the big flip is modeled as a single step (Fig. 2) preceded by a ProcedureInvocation that sets up the test CI; and, followed by a ProcedureInvocation that relocates the service and removes the old CI.

## V. TEST PLAN GENERATION

The goal of test plan generation is to design a test plan that enables the execution of TSIs under the required test configurations while maintaining the disturbance level within an acceptable range. Moreover, test plan designers may strive to reduce the disturbance induced and the time taken by testing activities to make such disturbance less noticeable or more tolerable.

Given a test suite and a set of test configurations against which each TSI is to be run, the number of the resulting test runs will always be the same for this combination regardless of how the test plan was designed. The cost we consider in this paper consists of the time taken and disturbance induced by the execution of a test plan and it can be broken down into: 1) a cost endured by running the TSIs; and 2) a cost endured by the setup and teardown of test configurations (setting up and removal of paths). Assuming that the former will be the same per TSI for all test plans involving a given test suite and a given set of test configurations. Thus, improvements can only be achieved by playing with the latter, i.e. the cost of setting up and tearing down test configurations. As a result, many activities that we propose in this approach focus mainly on reducing the number of times a test configuration is deployed, and its deployment time.

The test plan generation method is shown in Fig. 4. It starts by generating the test configurations under which each TSI is to be run. This generation is based mainly on the system configuration and the environment coverage criterion [13] the test plan designer provided as input for each TSI. Call path merging is an activity of the test plan generation that can be carried out while test configurations are being generated. This activity is the first and most important step in reducing the number of times test configurations are deployed. It mainly relies on the intersections of the call paths on which each TSI is to be applied, as well as the environment coverage associated with each TSI. Based on such information one can identify TSIs that will be associated with test configurations that can be deployed at once. The goal of call path merging is to put such TSIs into the same group. Therefore, there exists a test configuration, associated with a TSI of a group, that will have to be deployed for all the TSIs of the group to have their runs executed. After the call paths merging activity, the test method selection activity selects the test method that will be used for each CI in each call path associated with a group from the previous activity. After completing the test method selection and the test configuration generation are done, an initial UTP model is created using the mapping in Table I, and by cleaning up any duplicate test runs which may result from previous activities (mainly call paths merging). The initial UTP model is then given to the Test runs ordering activity which achieves two goals: 1) orders the test runs based on the precedence relationships between their associated TSIs; and, at the same time, 2) orders the test runs based on their associated test configurations to reduce disturbance. After the test runs ordering activity, the test plan generation is wrapped up by selecting the test runtime framework for each test component that will be involved in the test session.

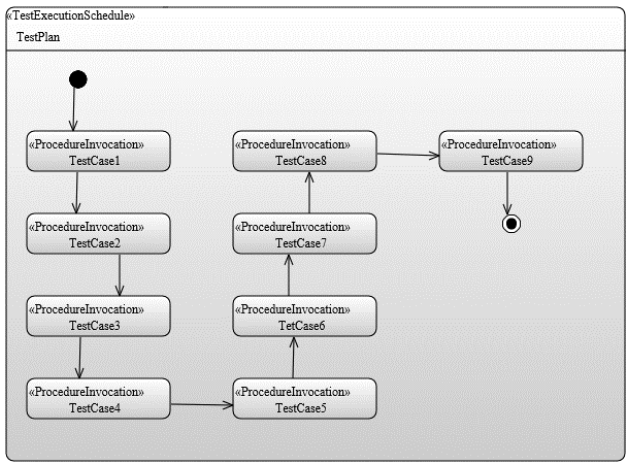

Fig. 3. Modeling of a rolling paths test method

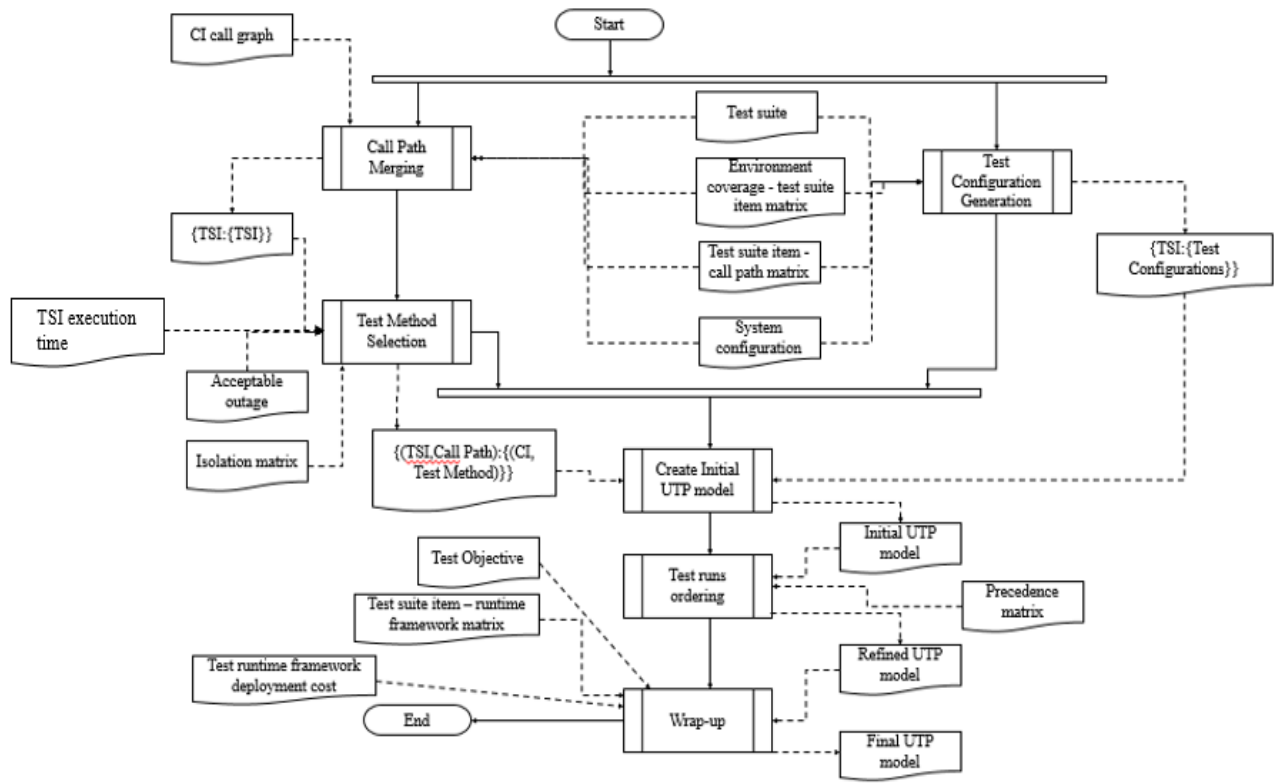

Fig. 4. Overall Test Plan Generation Approach

*A. Input artifacts*

The input artifacts of our method are as follow:

a) System configuration: composed of the configurations of all the CIs that compose the system.

b) Test suite: a set of test cases and test design techniques that will be used to achieve the test objective. Each element of the test suite is a TSI.

c) TSI application matrix: maps every TSI to a call path in the CIs call graph. The vertices of such a path represent the CIs that are targeted by the TSI.

d) Environment coverage-TSI matrix: maps every TSI to the environment coverage that its runs should achieve.

e) CIs call graph: a directed graph that captures functional dependencies between the CIs. Each vertex in this graph represents a CI in the system. An edge going from vertex V1 to vertex V2 means that the CI represented by V1 calls the CI represented by V2 in a realization of one of the services provided by the system. Each edge has a weight that represents the tolerance time of the CI represented by the source of the edge to the unavailability of the CI represented by the target of the edge. Such representation of the system can be extracted automatically using tools such as [14]. Moreover, the weights of the edges of such a graph (i.e. the tolerance time) is usually part of the configuration and indicates the outage unnoticeable for the dependent CI.

f) Isolation cost matrix: associates with every CI in the system the time it takes: 1) to snapshot one of its components, 2) to clone one of its components from an existing snapshot, and 3) to relocate from one of its components to another the portion of the SIs a component provides. Such information can be provided either by the CI vendor or measured when the CI is first acquired by the service provider.

g) TSI execution time: associates with each TSI an estimate of the time one of its runs may take. This information is usually provided by the test developer. In a context that leverages automation such as [1], such information can be collected from previous runs of the TSI or set by a default value.

h) Test runtime framework deployment cost: associates with every test runtime framework the deployment alternatives and the time it takes to deploy it using each possible alternative. Such information is usually available at test case (or TSI) development time.

i) TSI test runtime framework matrix: associates with every TSI the runtime framework that is needed to

execute the TSI. Such information is usually available at test case (or TSI) development time.

j) Acceptable outage: associates with every SI the duration for which it can be unavailable during the testing. A SI is said to be unavailable if it was inaccessible to a dependent SI for longer than the tolerance time of this dependent SI or if the availability manager reports it as unavailable (if the SI has no dependent). Note that acceptable outage is the time spent by the SI in an outage that is noticeable, while the tolerance time is the time for which the disturbance of the SI is tolerable.

k) Test objective: the objective that needs to be achieved by the test session for which the test plan is to be generated.

l) TSIs precedence matrix: maps each TSI in the test suite to TSIs in the test suite that have to be executed before it. Such information is usually available at TSI development time.

### B. Test configuration generation

The test configuration generation activity generates test configurations under which the TSIs will be run. In a test configuration, each CI along the call path is assigned a mixture, such an assignment specifies the path under which the TSI run is to be conducted. The test configuration generation takes as input the test suite (b), the environment coverage-TSI matrix (d), the system configuration (a), and the TSI-call path matrix (c). The generation of a test configuration is done with respect to the coverage criteria provided in (d) and they are amongst the ones defined in [13]. The generation of test configurations that takes into consideration the environment coverage has three main steps: First the boundary environments are identified from the configuration for the required coverage criterion. Then, the mixtures are created based on the set of boundary environments and the mixtures width. Finally, the set of test configurations that satisfy required criterion is created for the

- All boundary environment mixtures paths coverage: as the cartesian product of the sets of mixtures of the CI along the call path.
- Pairwise boundary environment mixtures: as the covering array of strength two created by considering each CI as a factor and each mixture of a CI is a level of the factor associated with that CI.
- All boundary environments mixtures: as the covering array of strength one created by considering each CI as a factor and each mixture of a CI is a level of the factor associated with that CI.

### C. Call paths merging

The call paths merging helps reduce the cost of testing by playing with the first factor that contributes to the cost which is the number of times test configurations are deployed. Because more than one TSIs may have runs under the same test configuration, deploying such test configurations only once and invoking the TSIs is indeed a way to reduce the number of test configurations deployments. This merging takes as input the test suite (b), the TSI call path matrix (c), the environment coverage-TSI matrix (d), and the CIs call graph (e). The output of the call paths merging activity is a set of groups of TSIs, the runs of TSIs of each group under a given test configuration are invoked within the same UTP TestCase in the final UTP TestExecutionSchedule model. A call path is a path (as defined in graph theory) in the CIs call graph (e). We say that path $A$ is a sub-path of path $B$ if: 1) all vertices of $A$ are also vertices of $B$; and 2) all edges of $A$ are also edges of $B$. One can also say that $B$ is a super-path to $A$. In $S$ set of paths, a max-path is a path which is a super-path to all paths in $S$.

The call paths merging follows two rules. A path $A$ can be merged with $S$ set of paths only if:

- $A$ is a super-path to the max-path of $S$, and the width of the mixtures in which $A$ is to be covered is greater than or equal to the maximum width in which the max-path of $S$ is to be covered.
  Or,
- $A$ is a sub-path of the max-path of $S$, and there exists at least one mixture width in which the max-path of $S$ is to be covered that is greater than or equal to the width of the mixtures in which $A$ is to be covered.

Applying these two rules may result in two types of merges:

- Full merge: which is a merge that happens when the sub-path has a weaker environment coverage criterion than the super-path. It is called full merge because the runs of the sub-path are covered by the runs of the super-path.
- Partial merge: which is a merge that happens when the sub-path has stronger environment coverage criterion than the super-path. It is called partial merge because the runs of the super-path will not be enough to cover all the runs of the sub-path. As a result, the runs of the sub-path may be split over several super-paths, and some runs may need to be covered in addition, and will be executed together.

**Algorithm 1:** Call Path Merging

```
1  CI_CG: (e),TS: Test Suite, EC_TSI: (d), TSI_CP: (c);
2  TSI_G: output = {};
3  while TS not Empty do
4      cTSI = TS.first();
5      if TSI_G not Empty then
6          for t in TSI_G do
7              if TSI_CP.get(cTSI).length == 1 then
8                  if TSI_CP.get(cTSI).isSubPath(TSI_CP.get(t)) then
9                      if EC_TSI.get(cTSI).isOfLessWidth(EC_TSI.get(t)) then
10                         TSI_G.get(t).add(cTSI); TS.remove(cTSI);break;
11                 if TSI_CP.get(cTSI).isSubPath(TSI_CP.get(t)) then
12                     if EC_TSI.get(cTSI).isWeakerThan(EC_TSI.get(t)) and EC_TSI.get(cTSI).isOfLessWidth(EC_TSI.get(t)) then
13                         TSI_G.get(t).add(cTSI);
14                         TS.remove(cTSI);break;
15                 if TSI_CP.get(t).isSubPath(TSI_CP.get(cTSI)) then
16                     if EC_TSI.get(t).isWeakerThan(EC_TSI.get(cTSI)) and EC_TSI.get(t).isOfLessWidth(EC_TSI.get(cTSI)) then
17                         tmp = TSI_G.get(t);
18                         tmp.add(cTSI);
19                         TSI_G.put(cTS,tmp);
20                         TSI_G.remove(t);
21                         AdjustGroupingFull(TSI_CP,EC_TSI,cTSI,TSI_G);
22                         AdjustGroupingPartial(TSI_CP,EC_TSI,cTSI,TSI_G);
23                         TS.remove(cTSI);break;
24             if TS.contains(cTSI) then
25                 isBeingMerged = false;
26                 for t in TSI_G do
27                     if TSI_CP.get(cTSI).isSubPath(TSI_CP.get(t)) then
28                         if (not EC_TSI.get(cTSI).isWeakerThan(EC_TSI.get(t))) and EC_TSI.get(cTSI).isOfLessWidth(EC_TSI.get(t)) then
29                             TSI_G.get(t).add(cTSI);
30                             if not isBeingMerged then
31                                 isBeingMerged = true;
32                     if TSI_CP.get(t).isSubPath(TSI_CP.get(cTSI)) then
33                         if (not EC_TSI.get(t).isWeakerThan(EC_TSI.get(cTSI))) and EC_TSI.get(t).isOfLessWidth(EC_TSI.get(cTSI)) then
34                             tmp = TSI_G.get(t);
35                             tmp.add(cTSI);
36                             TSI_G.put(cTSI,tmp);
37                             AdjustGroupingPartial(TSI_CP,EC_TSI,cTSI,TSI_G);
38                             TS.remove(cTSI);break;
39                 if isBeingMerged and TS.contains(cTSI) then
40                     TSI_G.put(cTSI,{cTSI});
41                     TS.remove(cTSI);
42             if TS.contains(cTSI) then
43                 TSI_G.put(cTSI,{cTSI});
44                 TS.remove(cTSI);
45     else
46         TSI_G.put(cTSI,{cTSI});
47         TS.remove(cTSI);
   procedure AdjustGroupingPartial(TSI_CP,EC_TSI,tsi,grouping)
       for nT in grouping do
           if TSI_CP.get(nT).isSubPath(TSI_CP.get(tsi)) then
               if not EC_TSI.get(nT).isWeakerThan(EC_TSI.get(tsi)) and EC_TSI.get(nT).isOfLessWidth(EC_TSI.get(tsi)) then
                   grouping.get(tsi).addAll(grouping.get(nT));
   procedure AdjustGroupingFull(TSI_CP,EC_TSI,tsi,grouping)
       for nT in grouping do
           if TSI_CP.get(nT).isSubPath(TSI_CP.get(tsi)) then
               if EC_TSI.get(nT).isWeakerThan(EC_TSI.get(tsi)) and EC_TSI.get(nT).isOfLessWidth(EC_TSI.get(tsi)) then
                   grouping.get(tsi).addAll(grouping.get(nT));
                   grouping.remove(nT);
```

The goal of the call paths merging activity is to perform as many full merges and partial merges as possible, thus reducing the number of times some test configurations will be setup to execute the TSIs runs.

Algorithm 1 achieves the goals of the call paths merging activity, i.e. it applies as many full and partial merges as possible. From the algorithm one can identify several possibilities of the merging. Lines 6-23 show the possible scenarios of the full merge. The full merge can either be a full merge while maintaining the same max-path (lines 6-14); or, a full merge in which the new TSI sets a new max-path for the group (lines 15-22). Similarly, partial merges are done in various forms (lines 24-42). The first scenario of a partial merge (line 24-31) consists of distributing the runs of a TSI over several groups with max-paths that are super-paths to the call path of the TSI. The algorithm accounts for the case where such groups are not enough to cover all the runs of the TSI, thus there is the addition of another group (Line 41) to cover the remaining runs. In the second scenario of the partial merge (lines 32-39), the max-path of the group to which the TSI is added is set by the newly added TSI. Therefore, this implies that all the runs of the new TSI are covered in this new group, but runs of the TSIs of the old group (the group before the addition of the new TSI) may not be all covered. Therefore, the algorithm keeps the old group as well to account for the runs that will not be covered by the group after the partial merge.

The goal of Algorithm 1 is to maintain the set of test configurations used during the execution of the test plan to the minimum. As a result, we aim to prove that the set of test configurations associated with the max-path of each grouping is the minimum set of test configurations needed to execute a test plan. One crucial observation in the construction of such a proof is that when the merging is done with a change of the max-path of the grouping the set of test configurations may increase, however, when a grouping in an iteration is done without change of a max-path the number of test configurations remains unchanged. The details of the proof can be found in Appendix A.

*D. Test method applicability and selection*

After grouping the test runs in the previous activity, we obtain sets of test runs that are grouped together. For each group of TSIs, the test method selection activity selects the test methods that will be used. Since the test methods apply at the CI level, for each group of TSIs we assign a test method for each CI in the max-path of that group of TSIs. The test method selection activity takes as input the groups of TSIs from the call paths merging and their associated paths, the system configuration (a), the call graph of the CIs (e), the isolation cost matrix (f), and the acceptable outage (j). The selection of the test methods takes into consideration the availability of resources, the cost of isolation, the dependencies between CIs, and the amount of tolerable disturbance for each SI.

*1) Applicability check of the test methods*

A test method is said to be applicable to a given CI if it can be used for the test isolation for this CI without causing any unacceptable outage. As a result, more than one test method may be applicable to a CI. Furthermore, a test method may be used even though it causes outage, but this outage is acceptable. The applicability of the different test methods determined as follows:

- The single step test method is applicable for CIs that do not present any potential risk of test interferences. In other words, the production traffic and test traffic can coexist without interference.

- The rolling paths test method is applicable to a CI if the time it takes to snapshot a component of that CI is less than the tolerance time of all the SIs that depend on the SIs provided by this CI and the service relocation time is also less than the tolerance time of all the dependents. I.e. dependent SIs will not notice neither the operation of taking a snapshot nor the operation of relocating the service.

- The small flip is applicable whenever the rolling paths is applicable, and when the configuration allows it. In other words, the small flip is applicable to a CI if there exists $S$, for which Equation (1) holds. In Equation (1) *env* is a boundary environment, $S$ is a set of boundary environments of the CI, $N$ is the width of the mixture of the environment coverage criteria, *Nodes(env)* is the set of nodes that can host boundary environment *env*, $K$ is the number of components needed by the CI at the time of the testing, and *Nodes* is the set of all nodes on which

$$\sum_{env \in S} \min(N, |Nodes(env)|) \leq |Nodes| - K - \left\lceil \frac{\sum TestConfSetupTime + \sum TestExecutionTime}{coolDownPeriod} \right\rceil \times scalingStep \quad (1)$$

the CI is deployed. Note that $K$, the number of components needed at the time of the testing, may not be known at the time of the design of the test plan. It can be estimated based on the operational profile of the CI, or its worst-case value can be used.

- If resources permit the big flip can always be used However, it is said to be applicable when the sum of the snapshot time, clone time, and the service relocation time is less than the acceptable outage. It is also preferred (in comparison with other methods) when the snapshot time or the service relocation time is more than the tolerance time of at least one dependent SI.

*2) Test method selection*

The test method selection is an important activity of the test plan generation, because the decisions made during this activity impact the disturbance induced by the test plan execution. On the one hand, the selected test method impacts the number of test runs that can be executed simultaneously which impacts the time it takes to execute the test plan. On the other hand, it impacts the number of service relocations for each CI which impacts the level of disruption induced by the test plan execution.

In order to keep the disruption acceptable and reduce the execution time, a test method can be selected for a given CI as follows:

- If only one test method is applicable (based on the applicability check), select that test method.

- If more than one test method is applicable, the precedence of test methods is single step, big flip, small flip, and last is the rolling paths.

```
Algorithm 2: Test method selection
1  TSI_G: Test suite items groupings;
2  Is_M: Isolation matrix;
3  A_O: Acceptable outage;
4  SysConf: System configuration;
5  CI_TM: Test method assignment;
6  foreach g in TSI_G do
7      Available_Resources = getAvailableResources(SysConf);
8      CI_TM.put(g.path,{});
9      for ci in g.path do
10         if getAplicableTM(Is_M,A_O,ci,SysConf).size()==1 then
11             CI_TM.get(g.path).add((ci,getAplicableTM(Is_M,A_O,ci,SysConf).first()));
12             updateAvailableResources(CI_TM,Available_Resource);
13     g.path.sortByMixtureSize();
14     for ci in g.path do
15         if getAplicableTM(Is_M,A_O,ci,SysConf).size()>1 and getAplicableTM(Is_M,A_O,ci,SysConf).size().contains(SingleStep) then
16             CI_TM.get(g.path).add((ci,SingeStep);
17             continue;
18         if getAplicableTM(Is_M,A_O,ci,SysConf).size()>1 and getAplicableTM(Is_M,A_O,ci,SysConf).size().contains(BigFlip) then
19             CI_TM.get(g.path).add((ci,BigFlip);
20             updateAvailableResources(
21             CI_TM,Available_Resource);
22             continue;
23         if getAplicableTM(Is_M,A_O,ci,SysConf).size()>1 and getAplicableTM(Is_M,A_O,ci,SysConf).size().contains(SmallFlip) then
24             CI_TM.get(g.path).add((ci,SmallFlip);
25             updateAvailableResources(
26             CI_TM,Available_Resource);
27             continue;
28         CI_TM.get(g.path).add((ci,RollingPaths);
29         updateAvailableResources(CI_TM,Available_Resource);
```

- Conflicts (e.g. due to resource needs) between two CIs are solved by setting the preferred test method for the CI with the bigger number of mixtures.

Algorithm 2 enables the test method selection according to these rules. For each TSI group it initializes the test method assignment to the empty set (Line 8). Then it assigns a test method to each CI to which only one test method is applicable (lines 9-12). It then sorts the remaining CIs in a decreasing order of their mixture size (i.e. number of mixtures) (Line 14). It iterates through this sorted set of CIs and from the applicable test methods of each CI it assigns the most preferred one (lines 16-17, 19-20, 23-24, 27) and updates the set of available resources as it is used in subsequent iterations to select the test methods for the remaining CIs. The getApplicableTM(…) function performs the applicability check to obtain the set of test methods that are applicable to a CI.

The test plan generation also aims at generating a test plan with a minimum execution time while maintaining the disturbance acceptable. In Algorithm 2 we follow two heuristics to achieve this goal. The first heuristic is the precedence order of the test methods. The second heuristic manages the conflicts when two CIs cannot be tested at the same time with their preferred test method. We evaluated Algorithm 2 to show to which extent our heuristics can help achieve the intended goal of the optimality of the execution time. We also listed the situations in which Algorithm 2 yields the optimal solution, and in which situations it does not as well as how bad the solution can be. The details of the evaluation can be found in Appendix B.

### E. Ordering of test runs

The ordering of test runs has two goals: 1) ensuring that TSIs are invoked only when their preconditions are met, and 2) reducing the disturbance by reducing the impact of service relocations. Each test run is a combination of a TSI and a test configuration under which it will be invoked. The first goal of the test runs ordering is achieved by ordering the test runs based on the TSIs they invoke. The second goal is achieved by ordering the test configurations in a way that the more critical a CI is, the least service relocations it experiences. The ordering of test runs takes as input a test plan with test runs in arbitrary order and the test suite precedence matrix (l). This process performs the re-ordering using the following operators on the test plan:

- Ordering test runs by changing the order of invocation of TSIs within the same UTP TestCase.

```
Algorithm 3: Test runs ordering
1  UTP_M: Initial UTP Model;
2  P_M: Precedence Matrix;
3  toBeMaintained = {};
4  foreach c in P_M do
5      if UTP_M.TestCases.select(t |t.invokes(c.preceding)).includesAll(UTP_M.TestCases.select(t |t.invokes(c.following))) then
6          for tc in UTP_M.TestCases.select(t |t.invokes(c.following)) do
7              placeAfter(tc,c.following,c.preceding);
8          readjust(UTP_M,toBeMaintained);
9          toBeMaintained.add(c);
10         continue;
11     if UTP_M.TestCases.select(t |t.invokes(c.following)).includesAll(UTP_M.TestCases.select(t |t.invokes(c.preceding))) then
12         for tc in UTP_M.TestCases.select(t |t.invokes(c.preceding)) do
13             placeAfter(tc,c.following,c.preceding);
14             for ftc in UTP_M.TestCases.select(t |t.invokes(c.following) and not t.invokes(c.preceding)) do
15                 placeAfter(UTP_M,ftc,tc);
16         readjust(UTP_M,toBeMaintained);
17         toBeMaintained.add(c);
18         continue;
19     if UTP_M.TestCases.select(t |t.invokes(c.following)).excludesAll(UTP_M.TestCases.select(t |t.invokes(c.preceding))) then
20         for tc in UTP_M.TestCases.select(t |t.invokes(c.preceding)) do
21             for ftc in UTP_M.TestCases.select(t |t.invokes(c.following)) do
22                 placeAfter(UTP_M,ftc,tc);
23         readjust(UTP_M,toBeMaintained);
24         toBeMaintained.add(c);
25         continue;
26 tcs = UTP_M.TestCases;
27 foreach tc in tcs do
28     toSort = UTP_M.TestCases.select(t |t.invokedTCs.forall(itc, tc.invokes(itc)));
29     currentTc = tc;
30     while toSort not Empty do
31         next = MaxSimilar(currentTc.conf,toSort);
32         placeAfter(UTP_M,currentTc,next);
33         toSort.remove(currentTc);
34         tcs.remove(currentTc);
35         currentTc = next;
```

- Ordering test runs by changing the order of UTP TestCases to order test runs based on the test configurations they involve.
- Ordering test runs by changing the UTP TestCase within which the TSI is invoked.

The first step of ordering test runs activity, described below, helps to achieve the first goal. After this step is completed, a TSI – the following TSI – will be invoked under a given test configuration only after all TSIs – the leading TSIs – that should precede this TSI (as per the precedence matrix) have been invoked under that same configuration. Therefore, this step uses only the first and third operators and proceeds according to the following rules:

- If the subset of UTP TestCases in which the leading TSI is invoked includes the subset of UTP TestCases in which its following TSI is invoked, then order the invocations of the TSIs within the same UTP TestCase of the subset of UTP TestCases in which the following TSI is invoked in a way that the following TSI is always invoked after the leading TSI.
- If the subset of UTP TestCases in which the following TSI is invoked is a union of a subset of UTP TestCases in which the leading TSI is invoked and a subset of UTP TestCases in which the leading TSI is not invoked; then for the first subset of the union order the invocations of the TSIs within the same UTP TestCase in a way that the following TSI is invoked after the leading TSI, and follow up this subset with the second subset of the union in which the leading TSI is not invoked.
- If none of the above rules apply, the third operator is used to move invocations of the following TSI to the first UTP TestCase in which it can be invoked (as per the test configuration of that UTP TestCase) while maintaining the precedence constraint. Moving such invocation may lead to the violation of Equation 1 in case any of the CIs in the max path is to be tested using small flip. When such violation occurs, the test method of such CI will change from small flip to rolling paths.

To achieve the second goal of the ordering, the test runs need to be ordered based on the test configurations they involve to reduce the disturbance induced by service relocations. The solution proposed in this paper is based on the following assumption: the more similar consecutive test configurations are, the less disruption services endure. To assess similarities between test configurations, we represent them as assignments of mixtures to nodes along a call path. Therefore, the similarity between two test configurations is the number of nodes along a call path to which different mixtures are assigned in the two test configurations. The ordering of UTP TestCases that takes into account such similarities goes along the following line:

- For each call path, start from a random UTP TestCase as the current UTP TestCase.
- The next UTP TestCase should be the one most similar to the current UTP TestCase, i.e. the UTP TestCase that involves a test configuration that changes the least number of mixtures from the test configuration of the current UTP TestCase. If more than one UTP TestCases change the same number of mixtures of the test configuration of the current UTP TestCase, the TestCase that changes the less critical CIs is chosen, i.e. disturbing less critical CIs is preferred compared to disturbing more critical CIs.

Algorithm 3 can be used to order test runs according to the rules mentioned above. It first achieves the first goal of the ordering of test runs by taking into consideration the precedence constraints (lines 4-31). Then it achieves the second goal of ordering of test runs by taking into consideration the test configurations the test runs involve (lines 33-43).

To achieve the first goal of test runs ordering, the first and third operators mentioned earlier are used. Lines 5-10 address situations in which the first operator is used to maintain precedence constraint by ordering TSI invocations within the same UTP TestCase. Lines 22-29 address situations in which the third operator is used in Line 25 to maintain the precedence constraints by ordering UTP TestCases in the UTP model. Lines 12-20 use both operators, first operator in Line 14, and third operator in lines 16 in order to maintain the precedence constraints. Note that every time a precedence constraint is handled it is added to a set of constraints, toBeMaintained, as readjustments with respect to these constraints are needed every time test runs are moved around to satisfy a new constraint.

To achieve the second goal of test runs ordering, i.e. reducing the disturbance, Algorithm 3 sorts UTP TestCases that invoke the same set of TSIs based on the test configurations they involve. To do so, for each UTP TestCase tc in the UTP model it starts first by finding the UTP TestCases that invoke the same set of TSIs as tc (Line 34). It places tc as the first UTP TestCase of that group, then places after tc the UTP TestCase that involves the configuration most similar to tc's configuration (lines 36-41). Every time a UTP TestCase is sorted it is removed from the set of UTP TestCases to be considered, this process keeps going until there are no more UTP TestCases to sort.

Algorithm 3 orders the test runs, first based on the precedence between the TSIs they involve and second based on the test configurations they involve. The second aspect of this ordering aims at reducing the disturbance that may be caused by moving from one test configuration to another. We prove in Appendix C that Algorithm 3 yields the optimal solution that reduces the disruption caused by going from one test configuration to the next.

### F. Wrapup

The wrapup activity helps completing the specification of the TestExecutinoSchedule. It takes as input the test objective (k), test runtime framework deployment cost (h), TSI-test runtime framework matrix (i), and the refined UTP model obtained from the test runs ordering activity. This activity starts first by adding the TestObjective to the UTP model, i.e. by creating the TestObjective model element and filling in its description attribute with the test objective given as input. Then it proceeds to choose the most suitable runtime framework deployment. This is done first by identifying the runtime framework of the TSI from (i), then checking the deployment options of this runtime framework in (h) (whether the runtime framework can be deployed using a configuration manager, or using a VM image, or a container). Then the least disturbing option is chosen, the order of precedence between the deployment options are container deployment, then VM deployment, then the deployment using a configuration manager when no other option is available.

## VI. AN ILLUSTRATIVE EXAMPLE

We implemented a prototype of our approach for test plan generation. The implementation was done using the Epsilon family of languages. Each one of the activities outlined in Fig. 5. is implemented as an Epsilon module. The test configuration generation, call path merging, test method selection, and the creation of initial UTP models are implemented using the Epsilon Object Language (EOL). The ordering of test runs is implemented using Epsilon Pattern Language (EPL). In this section, we will demonstrate the prototype through an illustrative example.

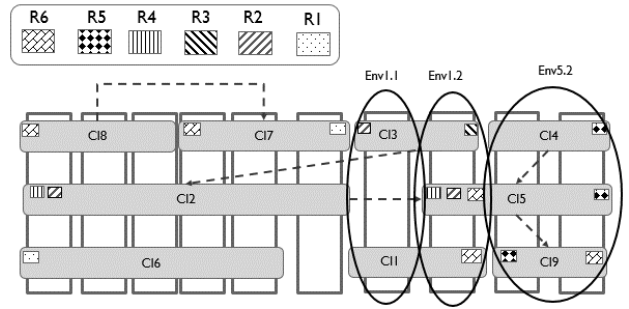

Fig. 5. System Configuration of the illustrative example

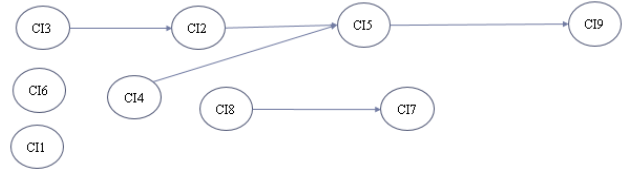

Fig. 6. ConfiguredInstance call graph associated with the configuration of Fig. 5.

The system configuration we take as input for this example is shown in Fig. 5. The system is composed of nine CIs, each one of them handling a certain number of SIs shown as smaller squares inside the big rectangle (four for CI5, two for CI2, CI3, CI7 and CI9, one for the rest of the CIs). The squares of SIs that contribute to the realization of the same requirement are shown with the same pattern as the requirement they help to realize, i.e. squares R1-R6. Some CIs depend on other CIs indicated by arrows (CI8 depends on CI7 for instance). From this configuration one can also identify the boundary environments to consider. The boundary environments related to CI1, for instance, are Env1.1 (which has a component of CI1 collocated with a component of CI3), and Env1.2 (which has a component of CI1 collocated with a component of CI3 and a component of CI5). CI3 has the same set of boundary environments as CI1. For CI5, in addition to Env1.2, it has also another boundary environment Env5.2 (which has a component of CI5 collocated with a component of CI4 and a component of CI9). The CI call graph associated with this configuration is shown in Fig. 6. An edge between two CIs indicates that the source CI provides at least one SI that depends on at least one SI of the target CI. It is taken as input in our prototype as an instance of the CI call graph metamodel. This metamodel allows capturing the concepts needed for a weighted directed graph.

The TSI call path matrix is show in Table II. We associate each TSI with the set of SIs it traverses; from this information and the CI call graph one can deduce to which call path each TSI applies. Certain TSIs apply to a single call path such as TC2 which applies only to the path CI3->CI2->CI5. Other TSIs may apply to more than one path such as TC1 which applies to CI8-

TABLE II. TSI CALL PATH MATRIX

| TSI | call paths |
| --- | --- |
| TC1 | {CI8->CI7, CI1} |
| TC2 | {CI3->CI2->CI5} |
| TC3 | {CI4->CI5->CI9} |
| TC4 | {CI2->CI5} |
| TC5 | {CI4,CI5} |

TABLE III. ISOLATION MATRIX (TIME UNIT: SECONDS)

| CI | Risk | Snapshot | Clone | Load relocation |
| --- | --- | --- | --- | --- |
| CI1 | 1 | 0.3 | 1 | 0.001 |
| CI2 | 1 | 0.7 | 1.2 | 0.01 |
| CI3 | 0 | 0.4 | 1.3 | 0.03 |
| CI4 | 0 | 10 | 10 | 5 |
| CI5 | 1 | 5 | 5 | 3 |
| CI6 | 0 | 10 | 9 | 13 |
| CI7 | 1 | 4 | 5 | 3 |
| CI8 | 0 | 1 | 1 | 1 |
| CI9 | 1 | 0.1 | 0.1 | 0.1 |

>CI7 and CI1. Such differences may arise when some TSIs aim to validate the service of a specific tenant and which realize a certain requirement (the case of TC2), while others aim to validate the realizations of a specific requirement for more than one tenant (the case of TC1). As a result, we need to append indexes to the TSI Ids to remove ambiguity (for instance TC1-0 is the application of TC1 to CI8->CI7 and TC1-1 is the application of TC1 to CI1).

In this case study we consider a simple environment coverage case. The coverage criterion is the same for all the TSIs and it is the all boundary environment mixtures coverage. Moreover, we consider the case of mixtures of width one. As a result, the set of test configurations generated for each TSI should involve each mixture of width one (i.e. boundary environment) of each CI along the call path at least once. Table III shows an example of the isolation matrix. In this matrix, for each CI one can find in the first column whether the CI represents a risk of interferences (1 means there is a risk of interferences while 0 means there is no risk of interferences), and, in the rest of the columns respectively, the time needed for snapshotting, cloning, and relocating the service. This information along with the acceptable outage guides the choice of the test method for each CI.

Fig. 7. shows the results of running the test plan generation prototype. The model elements that are first created in the model are the OpaqueBehavior model elements. These elements can be either for deploying or for removing test configurations (non-stereotyped OpaqueBehavior model elements); or, for invoking TSIs in which case they will be stereotyped with the UTP stereotype TestProcedure. Taking into consideration the test runs needed for the test session, UTP TestCase model elements are created (as Activity elements stereotyped with TestCase). Each TestCase model element invokes other model elements in the following order: first, an OpaqueBehavior as setup (to deploy the test configuration), then an OpaqueBehavior stereotyped as TestProcedure as its main procedure to invoke the TSIs; and, finaly it invokes an OpaqueBehavior as teardown to remove the test configuration. After the creation of TestCases, an Activity stereotyped as TestExecutionSchedule is created, and it invokes the created TestCases taking into account all the ordering constraints to respect. Fig. 7(a) shows an OpaqueBehavior used to deploy a test configuration, which can be seen in the Body attribute of the OpaqueBehavior, i.e. "deploy {CI3:{E3.1},CI2:{E2.1},CI5:{E1.2}}". One can also notice how the groups manifest at the level of OpaqueBehavior elements that are seteretyped by TestProcedure. Fig. 7(b) shows a TestProcedure that invokes the grouped TSIs TC5-1, TC4-0, and TC2-0. Activities stereotyped as TestCase are then created as show in Fig. 7(c) to capture the execution of a set of grouped TSIs under a specific test configuration. This grouping is done using CallBehaviorAction elements which invoke OpaqueBehaviors such as the ones shown in Fig. 7(a) and Fig. 7(b). Each invoked behavior has a role (setup, main, or teardown), taking into consideration that UTP only allows TestProcedures to be invoked as main. Finally, the TestCase elements are invoked within an activity stereotyped TestExecutionSchedule as shown in Fig. 7(d).

VII. CONCLUSION

Test planning is an important task for the orchestration of test activities in production. Inappropriate test planning may lead to an orchestration that induces some unnecessary outages or can even lead to the violation of some functional or non-functional requirements of the system. Designing test plans manually is complex, tedious and error prone. In this paper, we proposed an approach to automate the generation of test plans. The proposed approach uses UTP as the modeling framework for test plans and uses the test methods proposed in [13] to allow for safe orchestration of test cases in production. Our proposed solution relies/acts on UML and UTP concepts at a high level of abstraction. Our algorithms are expressed as model transformations; thus, decoupling the details of our proposed solution from the specifics of each system and platform. In other words, our approach would apply to cloud services whether they involve container-based virtualization or VM based virtualization. Moreover, the expressiveness of UML and UTP enabled dealing with the diversity of technologies often encountered in cloud systems. The implementation of our approach is done using the Epsilon family of languages in the Papyrus environment. Each one of the activities outlined in Fig. 4. is implemented as an Epsilon module. The test configuration generation, call path merging, test method selection, and the creation of initial UTP models are implemented using the Epsilon Object Language (EOL). The ordering of test runs is implemented using Epsilon Pattern Language (EPL). The use of EPL eased the implementation of test runs ordering based on the TSIs they invoke. In fact, such ordering can be expressed as a single pattern that EPL finds in the UTP model and applies the appropriate operator to perform the ordering. EPL also offers the possibility of running the transformation on the model as long as the pattern can still be found, which helps maintain the previously processed TSIs in the appropriate order. Under the assumption that there are no cyclic dependencies between TSIs, which is always the case, such transformations terminate.

To the best of our knowledge, our approach is the first to address the problem of test isolation as part of the automation of test planning. In other words, test interferences are considered as a risk when planning testing activities in the production environment, and our approach proposes how such risk can be handled as part of the risk handling that takes place during test planning. Evaluation of the algorithms involved in the approach was performed as well. We have showed that our algorithm for call paths merging yields the minimum set of test configurations that are needed to execute the test plan. We have also shown that our test runs ordering algorithm minimizes the disturbance that may be endured when moving from one test configuration to the next. Our algorithm for test method selection was evaluated and we identified situations in which the solution can be optimal and when it is not. The algorithms in our solution address situations in which only one test plan is being executed in the production environment, and during such execution only one max call path is tested at a time. In other words, all the TSIs invoked between the deployment and removal of a given test configurations are applied to call paths with the same max call path. Discarding such constraint and allowing for TSIs that involve call paths with different max call paths, such as the execution of test runs of two different groupings simultaneously, may offer the opportunity to further optimize the test plan. However, the increase in the complexity of the problem is to be investigated. Furthermore, our observation regarding the impact that the effort needed to deploy/remove test configurations has on the execution time of a test plan can be further exploited to devise better solutions for test method selection. In fact, the validity of this observation does not depend on the test methods used in our solution nor on our definition of test configurations, thus making it a good basis for future solutions that establish test isolation policies or automate test planning in other contexts.

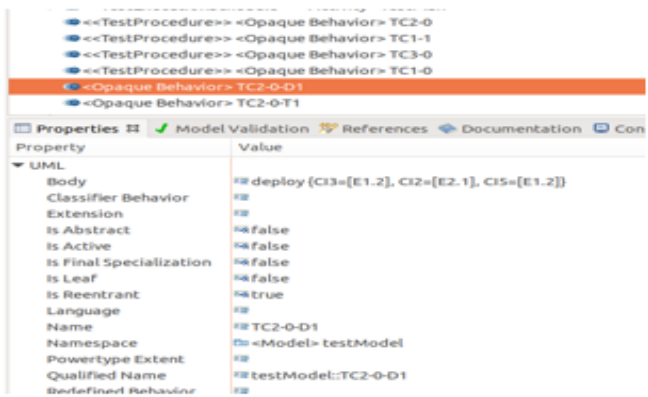

(a)

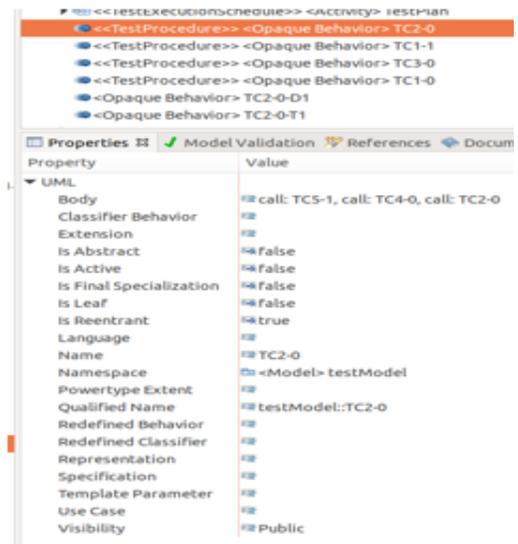

(b)

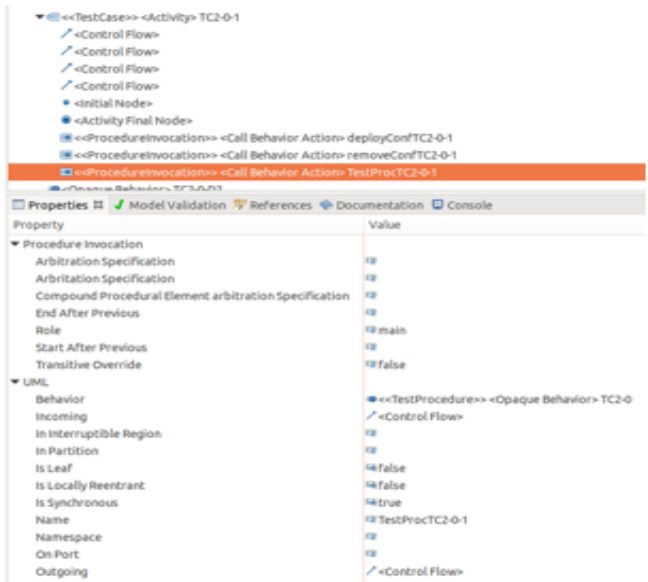

(c)

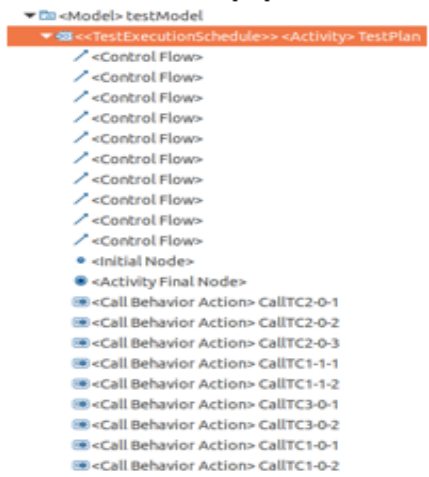

(d)

Fig. 7. Output UTP Model

ACKNOWLEDGMENT

This work has been partially supported by Natural Sciences and Engineering Research Council of Canada (NSERC) and Ericsson.

REFERENCES

[1] O. Jebbar, F. Khendek, M. Toeroe, Architecture for the Automation of Live Testing of Cloud Systems, in the proceedings of the 20th IEEE International Conference on Software Quality, Reliability, and Security, IEEE QRS '2020.

[2] M. Elqortobi, J. Bentahar, R. Dssouli, Framework for Dynamic Web Services Composition Guided by Live Testing, in the proc. Of Emerging Technologies for Developing Countries, AFRICATEK, 2017. Lecture Notes of the Institute for Computer Sciences, Social Informatics and Telecommunications Engineering, vol 206. Springer, Cham.

[3] A. G. Sanchez, Cost Optimizations in Runtime Testing and Diagnosis, PhD Thesis, Delft University of Technology, September 2011.

[4] C. Tang, T. Kooburat, P. Venkatachalam, A. Chander, Z. Wen, A. Narayanan, P. Dowell, R. Karl, Holistic Configuration Management at Facebook, in the proceedings of the 25th Symposium on Operating Systems Principles, SOSP '2015.

[5] Object Management Group. UML Testing Profile 2 (UTP2) Version 2.1

[6] S. Dathathraya, M. Stevens, System and method for generating automatic test plans, U.S Patents: US20050172270A1.

[7] E. B. Lewis, Technique for automatically generating a software test plan, U.S patents: US6546506B1.

[8] J. G. Becker, K. L. McClamroch, V. Raghavan, P. Sun, Automated test execution plan generation, U.S patents: US8423962B2.

[9] M. Jibbe, Method and system for generating a global test plan and identifying test requirements in a storage system environment, U.S patents: US20070079189A1.

[10] ISO/IEC/IEEE 29119-1, Software and systems engineering – Software testing – Part 1: Concepts and definitions, First edition, 2013.

[11] ISO/IEC/IEEE 29119-2, Software and systems engineering – Software testing – Part 2: Test processes, First edition, 2013.

[12] M. Lahami, M. Krichen, M. Jmaiel, Safe and efficient runtime testing framework applied in dynamic and distributed systems, Science of Computer Programming, Vol. 122, 2016.

[13] O. Jebbar, F. Khendek, M. Toeroe, Methods for Live Testing of Cloud Services. in the proceedings of the 32nd IFIP International Conference on Testing Software and Systems, ICTSS, 2020.

[14] Weave scope. https://www.weave.works/oss/scope/

APPENDIX A

The goal of the proof is to demonstrate that the call paths merging using Algorithm 1 yields the maximum merging, i.e. the set of test configurations associated with the groupings is the minimum set of test configurations to be deployed for all the test runs to be executed. To prove this, we define a partial order amongst test runs based on the test configurations they involve. We represent this binary relation as a directed graph and show afterwards that Algorithm 1 yields the sources of this graph as heads of the groupings. The number of sources is equal to the minimum number of test configurations to be deployed for all the test runs to be executed.

**Definition 1**: Let $TestConf1$ and $TestConf2$ be two test configurations, we say that $TestConf1$ is a sub-configuration of $TestConf2$ and write it as $TestConf1 < TestConf2$ iff:
$$\forall CUT \in TestConf1, CUT \in TestConf2 \land$$
$$env(CUT, TestConf1) = env(CUT, TestConf2)$$

Fig. 8. shows a representation of the partial order of the test runs according to the definition. The colors of the nodes distinguish the test cases involved in each test run, a test case with red nodes, a test case with black nodes, and a test case with grey nodes. Each node represents a run of the test case against a specific test configuration. An edge between two nodes indicates that the test configuration of the target node is a sub-configuration, as per the Definition 1, of the test configuration of the source node. (a) shows the case in which we may end up with a full merge, as every test configuration under which the test case with the black nodes is to be run is a sub-configuration of at least one test configuration under which the test case with red nodes is to be run. In such case the grouping will group the test runs of both test cases and the head of the grouping will be the test case with the red nodes. (b) captures two test cases that are run under unrelated test configurations, in this case the minimum number of test configurations that need to be deployed for all test runs to be executed equals the total number of test

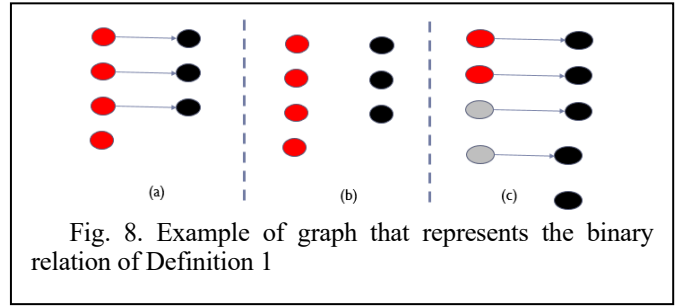

Fig. 8. Example of graph that represents the binary relation of Definition 1

configurations as no merging is possible. (c) shows the case in which we may end up with two partial merges as some test configurations under which the black test case is to be run are sub-configurations of test configurations under which the red or grey test case, moreover there are test configurations under which the black test case is to be run and which are not sub-configurations of any other test configuration. Such case occurs when the environment coverage under which the black test case is to be run is stronger than the environment coverages under which the red test case and the grey test case are to be run. In this situation we end up with three groupings, the first one has the red test case as head, the second one has the grey test case as head, and the third has the black test case as head in order to cover the test configurations that are not related to any other test configuration.

**Lemma 1:** given a test case $T$ and a set of test configurations $TConfs$ under which $T$ is to be run. The test configurations in $TConfs$ are un-ordered as per Definition 1.

In order to prove the optimality of Algorithm 1, we will proceed by induction:

- Case we have only one test case to be run under a set of test configurations. In this case Algorithm 1

will yield one grouping under which there is one test case, and as per Lemma 1, the number of test configurations to be deployed for the test runs to be executed will be equal to the number of sources in the graph representation of the ordering defined in Definition 1 or this set of test runs.

- Let $TCs$ be a set of test cases, each of which is to be run under a set of test configurations $TConfs(TC_i)$ for $1 \leq i \leq |TCs|$. Let $G_n$ be the result of Algorithm 1 when applied to the test cases in $TCs$ and their corresponding test configurations. Let $OR_n$ be the directed graph representation of the ordering defined in Definition 1 for the test runs obtained by running the test cases in $TCs$ under their corresponding test configurations. Let's assume that $G_n$ is optimal, i.e. the number of test configurations to be deployed for the test runs to be executed given $G_n$ equals the number of sources in $OR_n$. Let $TC_{n+1}$ be a new test case that is to be executed under a set of test configurations $TConfs(TC_{n+1})$, and let $OR_{n+1}$ be the directed graph representation of the ordering defined in Definition 1 for the test runs obtained by running the test cases in $TCs \cup \{TC_{n+1}\}$ against their corresponding test configurations. $OR_{n+1}$ will be composed of the nodes and edges of $OR_n$ plus additional nodes and egdes to represent the new test runs and their relation to the test runs from $TCs$. Since $OR_{n+1}$ and $OR_n$ are directed graphs, there are few ways into which such nodes and edges can be added:

    o Case 1: all the nodes that were added have no incoming edges from or outgoing edges to any of the sources of $OR_n$, i.e. new nodes are added in a manner like Fig. 8. (b). In this case $OR_{n+1}$ will have $|TConfs(TC_{n+1})|$ more sources than the sources of $OR_n$. Moreover, line 46 to line 47 of Algorithm 1 will be executed, in which case a new grouping will be added in which $TC_{n+1}$ will be the head, meaning that executing the test runs given the new grouping will result in the deployment of $|TConfs(TC_{n+1})|$ more test configurations, which is equal to the number of sources in $OR_{n+1}$. As a result, the solution is still optimal.

    o Case 2: a subset of the nodes that were added has outgoing edges to sources in $OR_n$, and such sources correspond to the test runs of a set of heads of groupings in $G_n$, i.e. the case in Fig. 8. (a) in which the nodes of the black test case existed in $G_n$ and $TC_{n+1}$ was the red test case. If $TC_j$ for $1 \leq k \leq j \leq l \leq |TCs|$ is the set of heads of groupings in question, then $OR_{n+1}$ has $|TConfs(TC_{n+1})| - \sum_{j=k}^{l}|TConfs(TC_j)|$ more sources than $OR_n$. Algorithm 1 will apply line 19 to apply a full merge when encountering the first head of such set, then loops through existing groupings in lines 21 to 22 to see if there is potential for any other full merge or partial merge. Therefore, the new solution will add $|TConfs(TC_{n+1})|$ (line 19) more test configuration deployment and substract $\sum_{j=k}^{l}|TConfs(TC_j)|$ (lines 21-22) test configuration deployments, which corresponds to the number of sources in $OR_{n+1}$. As a result, the solution is still optimal.

    o Case 3: the set of nodes that were added has outgoing edges to sources in $OR_n$, and such sources correspond to subsets of test runs of a set of heads of groupings in $G_n$, i.e. the case in Fig. 8. (c) in which the nodes of the black test case and the red test case existed in $G_n$ and $TC_{n+1}$ was the grey test case. If $TC_j$ for $1 \leq k \leq j \leq l \leq |TCs|$ is the set of heads of groupings in question, then $OR_{n+1}$ has the same number of sources as $OR_n$. Algorithm 1 will apply line 36 to apply a partial merge when encountering the first head of such set, then loops through existing groupings in line 37 to see if there is potential for any other partial merges. Since it is a subset and not the full set of test runs that existed in $OR_n$, this will not change the number of sources although it does change the heads of some groupings. As a result, the solution is still optimal.

    o Case 4: the set of nodes that were added has no outgoing edges to sources in $OR_n$, but has incoming edges from sources in $OR_n$, i.e. a manner like Fig. 8. (a) in which $TC_{n+1}$ is the black test case, or Fig. 8. (c) in which $TC_{n+1}$ is the black test case without the run represented by the extra node with no incoming or outgoing edges. In such situation $OR_{n+1}$ has the same number of sources as $OR_n$. Algorithm 1 will either execute lines 6 to 14 to perform a full merge (same situation as Fig. 8. (a)) or lines 27 to 31 to perform a set of partial merges (same situation as Fig. 8. (c)). In the latter case Algorithm 1 will also add an extra grouping (line 40), but such grouping will be filtered out in the wrap up as it will not contain any test run. As a result, the solution is still optimal.

    o Case 5: a subset of the nodes that were added has incoming edges from sources in $OR_n$ and another subset has outgoing edges to sources in $OR_n$. Let $I_n$ be the set

of test cases of which the runs are represented by sources that have outgoing edges to nodes that represent the runs of $TC_{n+1}$, and $O_n$ the set of test cases of which the runs are represented by sources that have incoming edges from the new sources that represent the runs of $TC_{n+1}$. $I_n$ and $O_n$ are both non-empty sets because if at least one of them is empty we end up in one of the cases 1-4. Obviously, the number of nodes that represent the runs of $TC_{n+1}$ is at least as big as the number of sources corresponding to test runs of the test cases in $I_n$ and $O_n$. If those numbers are equal, then the number of sources in $OR_{n+1}$ is equal to the number of sources in $OR_n$. If those numbers are not equal then $OR_{n+1}$ has $|TConfs(TC_{n+1})| - |\{ s \in sources(OR_n) \text{ such that } TSI(s) \in I_n \cup O_n\}|$. Using Definition 1, one can deduce that in the process of creating $OR_n$ the test runs of the test cases in $O_n$ were merged in a partial merge with the runs of the test cases in $I_n$. As a result, Algorithm 1 will use lines 27 to 31 to perform a partial merge with the nodes that correspond to test cases in $I_n$ without head change, and lines 32 to 38 to perform the partial merge with head change with nodes that correspond to test cases in $O_n$. Therefore, new sources will be considered only when the merge with the nodes in $O_n$ is a full merge, and it will be the case in which new sources are added to compose the solution. Hence the optimality of the result.

As a conclusion, Algorithm 1 yields the optimal solution for selecting the minimum set of test configurations to be used for all the test runs to be executed.

APPENDIX B

The goal of test plan generation is to design a test a plan that has a reduced execution time while maintaining the disturbance acceptable. The test configurations deployment effort is the only aspect on which a test plan designer can act in order to achieve this objective. The precedence order among test methods and the strategy that we have adopted in our algorithm to solve conflicts, both compose the set of actions taken in our test method selection in order to reduce the test configurations deployment effort. Such reduction manifests as a reduced number of instantiations/removals of the components under test; and, a leveraged use of parallelism as compared to other alternatives for designing the test plan. As a result, the goal of this evaluation is to answer two research questions:

- **RQ1**: Does the precedence order we use in our test method selection help reduce the test configurations deployment time?
- **RQ2**: How good is the solution obtained by using our strategy for conflict resolution?

Table IV shows the number of instantiations, removals, and service relocations needed to deploy the test configurations using each of the test methods we have proposed, these numbers are given in terms of the number of iterations and the number of components under test (CUTs) needed to deploy the test configurations. Table V gives the formulas for the number of iterations in terms of the number of test configurations when each test method is used. Finally, Table IV gives the number of test configurations depending on the coverage criteria associated with each TSI.

In our test method selection algorithm, when more than one test method is applicable, we give precedence for big flip, then small flip, and rolling paths is the last choice. From Tables IV, Table V and Table VI we see that the big flip and the small flip have the same number of instantiations, removals, and service relocations while the rolling paths has more instantiations and removals than the other two test methods. Moreover, knowing that the big flip allows for more parallelism than the small flip, as seen from Table III, we can deduce that our precedence order helps reduce the test configurations deployment time which answers RQ1.

When two CIs can be tested using more than one test methods, but only one of them at a time can have their preferred test method, our algorithm assigns the preferred test method of the CI with the biggest number of mixtures. To answer RQ2, we will go case by case (based on the environment coverage used) and evaluate how good the resulting solution is, we will show optimality in certain cases, give upper bounds for how bad a solution can be in other cases whenever possible. To achieve this goal, we will reason only on cases where the conflict is between the big flip and small flip. In fact, the big flip and small flip have the same number of instantiations and removals as we can see from Table IV, and this is the only relevant factor for test plan's execution time. Therefore, choosing one or the other does not make a difference in execution time as much as it does in resource consumption or disruption. Moreover, the results that will hold for one will also hold for the other. In the rest of this proof, we are reasoning only on a single instance of the test method selection problem. In other words, we are not addressing the decisions made for the whole test plan, we are focusing on the decisions made only for a single call path, i.e. one grouping. This is reasonable as the test method selection for one call path is totally independent of the test method selection for another call path as the testing in our test plans targets one call path at a time.

TABLE IV. NUMBER OF INSTANTIATION, REMOVAL, AND SERVICE RELOCATION PER TEST METHOD

| Test method | Number of instantiations | Number of service relocations | Number of removals |
|---|---|---|---|
| Big flip | $\#CUTs$ | $iterationCount$ | $\#CUTs$ |
| Small flip | $\#CUTs$ | $iterationCount$ | $\#CUTs$ |
| Rolling paths | $iterationCount$ | $iterationCount$ | $iterationCount$ |

TABLE V. ITERATION COUNT PER TEST METHOD IN TERMS OF THE NUMBER OF TEST CONFIGURATIONS

| Test method | iterationCount |
|---|---|
| Big flip | $\dfrac{\#testConfigurations}{min_{i,\ CI\ in\ call\ path}(\frac{\#comp_i}{mw_i})}$ |
| Rolling paths | $\#testConfigurations$ |
| Small flip | $\dfrac{\#testConfigurationsOfFirstBatch}{min_{i,\ CI\ in\ call\ path}(\frac{\#comp_i}{mw_i})} + \dfrac{\#testconfigurations - \#testConfigurationsOfFirstBatch}{min_{i,\ CI\ in\ call\ path}(\frac{\#comp_i}{mw_i})}$ |

Let $G$ be the number of instantiations/removals associated with the CIs being tested using the rolling paths test method in our solution. Let $OPT$ be the number of instantiations associated with the CIs being tested using the rolling paths test method in the optimal solution.

**Case 1: all boundary environment mixtures**

In this case $G$ and $OPT$ are given by the formula in the first row second column in Table VI considering only the set of CIs being tested using the rolling paths test method and not all the CIs in the call path. We will use an exchange argument to show that our solution is optimal in this case.

Let's assume the solution obtained by our algorithm is not optimal and see what kind of change we can perform to improve it. Let $CI_i$ be a CI that will be tested using the big flip in our solution, and $CI_j$ a CI that will be tested using the rolling paths in our solution. The only situation in which we can improve our solution by changing the test methods of $CI_i$ and $CI_j$ is if:

- $CI_j$ is the CI with the maximum number of mixtures amongst all the CIs being tested using the rolling paths in our solution.
- $CI_i$ has a number of mixtures less than the number of mixtures of $CI_j$.
- There are enough resources to test using the big flip $CI_j$ and all the CIs that have their number of mixtures bigger than the number of mixtures of $CI_j$.

Because of the third point such exchange is not possible. If the third condition was met, the big flip would have been chosen for $CI_j$ as per our algorithm. As a result, any exchange can only make our solution worse. Therefore, our solution is optimal.

**Case 2: all boundary environment mixtures paths**

In this case $G$ and $OPT$ are given by the formula in the first row first column in Table VI considering only the set of CIs being tested using the rolling paths test method.

**Lemma 2**: given a situation when test method selection is to be performed, the number of CIs tested using the big flip in our solution is at most equals the number of CIs tested using the rolling paths in the optimal solution.

Proof: we will go case by case:

TABLE VI. NUMBER OF TEST CONFIGURATIONS BASED ON THE ENVIRONMENT COVERAGE THAT WAS USED

| | All boundary environments mixtures paths | All boundary environment mixtures |
|---|---|---|
| $\#testConfigurations$ | $\prod_{i,\ CI\ in\ call\ path} C^{mw_i}_{\sum_{i \in BEs} \min(\#BE_i, mw_i)}$ | $max_{i,\ CI\ in\ call\ path}(C^{mw_i}_{\sum_{i \in BEs} \min(\#BE_i, mw_i)})$ |
| $\#testConfigurationsOfFirstBatch$ | $\prod_{i,\ CI\ in\ call\ path} C^{mw_i}_{\sum_{i \in S} \min(\#BE_i, mw_i)}$ | $max_{i,\ CI\ in\ call\ path}(C^{mw_i}_{\sum_{i \in S} \min(\#BE_i, mw_i)})$ |

- Case 1: for all the CIs for which a test method is to be chosen, the optimal solution and our solution use different test methods. In this case the number of CIs tested using the big flip in our test method equals the number of CIs tested using the rolling paths in the optimal solution.
- Case 2: for some CIs, our solution and the optimal solution use the same test method. If we assume that the number of CIs being tested using the big flip in our test method is more than the number of CIs being tested using the rolling paths test method in the optimal solution, then to improve our solution we need to test more CIs using the rolling paths. This is not possible since our heuristic favors the big flip for CIs with bigger number of mixtures. As a result, using the rolling paths more, can only make our solution worse.

**Definition 2**: let $CI_1$, $CI_2$, ..., $CI_n$ be a set of CIs with their associated numbers of mixtures $\#mx_1$, $\#mx_2$, ..., $\#mx_n$. We define $K$ such that: $\forall l \geq K, \forall i \in [|1, n|], \forall (\alpha_j)_{1 \leq j \leq l} \in ([|1, n|]\setminus\{i\})^l$, $\#mx_i \leq \prod_{j=1}^{l} \#mx_{\alpha_j}$

In other words, $K$ is the lowest number bigger than the maximum number of CIs that you can switch from the rolling paths to the big flip in exchange of one CI from the big flip to the rolling paths and improve $G$.

Let $R$ be the number of CIs for which the rolling paths test method was chosen in our solution. Let $B$ be the number of CIs for which the big flip was chosen in our solution. And let $l$ be the number of CIs for which the rolling paths test method was chosen in the optimal solution. In the case when $l > K$, if we choose for each CI being tested using the rolling paths in our solution, $K$ CIs from the ones being tested using the rolling paths in the optimal solution, and apply the inequality in Definition 2 while trying to diversify the CIs picked from the optimal solution as much as possible (using permutations), we can establish that:

$$G \leq OPT^{\left\lceil \frac{R \times K}{l} \right\rceil}$$

Since we cannot know $l$ without calculating the optimal solution, we will use $B$ and Lemma 2 to identify when this inequality is satisfied. Table VII captures the different upper bounds we can establish for our solution compared to the optimal solution. It is not obvious from the table but for $K = 2$ our solution is optimal. In fact, this can be proven using an exchange argument same as we did for the case of all boundary mixtures coverage. In some other cases one can obtain a solution that is no worse than $OPT^2$. In fact, if each CI in the call path cannot be collocated with any other CI in the call path, algorithms that are used for 2-approximations for knapsack problem will yield such solutions.

TABLE VII. UPPER BOUNDS OF OUR SOLUTION COMPARED TO THE OPTIMAL SOLUTION IN THE CASE OF ALL BOUNDARY ENVIRONMENT MIXTURES PATHS ENVIRONMENT COVERAGE

| Case | Upper bound |
|---|---|
| $R > B \geq K$ | $OPT^{K^2}$ |
| $R < B \wedge B \geq K$ | $OPT^K$ |
| Other cases | The solution can be arbitrarily bad |

APPENDIX C

The goal of test runs ordering based on the test configurations they involve is to reduce disturbance. This is done by reducing the number of service relocations configured instances endure. Moreover, it aims to maintain that more critical configured instances undergo less service relocations than less critical CIs. The goal of this proof is to demonstrate that the ordering obtained using Algorithm 3 induces minimum disturbance when used to execute TSIs. We also show that the more critical a CI is the less disturbance it endures when the ordering obtained using Algorithm 3.

**Lemma 3:** To move from one test configuration to another, it takes a service relocation of at least one CI being tested using rolling paths.

**Lemma 4:** Along a call path CIs being tested using big/small flip undergo one service relocation, CIs being tested using single step undergo zero service relocation, CIs being tested using rolling paths undergo a number of service relocations that depends on the ordering of test configurations.

In our proof we will first assume that we are testing call path along which all CIs are to be tested using the rolling paths (because of **Lemma 4**), then we will generalize the result. To perform the first step, we will go case by case depending on the environment coverage criterion that was used:

- **Case#1: all boundary environments mixtures coverage**
  - When using this coverage criterion, each CI goes through at least its $\#BMxs$ (number of boundary environment mixtures).
  - The generation of test configurations is done by varying as many CIs as possible (the greedy algorithm mentioned in the test configuration generation subsection), until no CI has more mixtures to cover. As a result, the later the test configurations are generated the more similar they are.
  - The number of test configurations generated is max ($\#BMxs$) as per Table IV
  - As a result, the algorithm will order the test configurations in a reverse order of their generation. And
  - each CI will go through exactly $\#BMxs$ of service relocations which is the minimum.
- **Case#2: all boundary environments mixtures paths coverage**

- The set of test configuration is the Cartesian product of the sets of boundary environment mixtures of the CIs along the path.
- The algorithm we propose when used is equivalent to the following mapping:
    - Each CI is represented by a digit.
    - The most critical CI is the most significant digit (the one on the extreme left), the least critical CI is the least significant digit (the one on the extreme right).
    - Each digit takes values in the range of $\#BMxs$ of its associated CI.
- The output of the algorithm is the n-ary Gray code associated with this setting
- Gray code is a single distance code.
- As a result, going from one test configuration to another is done using a single service relocation (the minimum as per **Lemma 3**).
- Moreover, because of how the CIs are prioritized (i.e. associated to digits).
- The least critical CI will undergo $\frac{\#testconfigurations}{\#BMxs}$ service relocations.
- The most critical CI will undergo its associated $\#BMxs$ of service relocations.